\documentclass{article}
\usepackage[utf8]{inputenc}
\usepackage{spconf,amsmath,graphicx}
\usepackage{booktabs}
\usepackage{url}
\usepackage{graphicx} % Make sure this is in your preamble
\usepackage{multirow, booktabs}

\title{CL-UZH submission to the NIST SRE 2024 Speaker Recognition Evaluation}

\name{%
  \parbox{\linewidth}{\centering
    Aref Farhadipour$^{1}$, Shiran Liu$^{2}$, Masoumeh Chapariniya$^{1}$, Valeriia Vyshnevetska$^{1}$,\\
    Srikanth Madikeri$^{1}$, Teodora Vukovic$^{1}$, Volker Dellwo$^{1}$
  }%
}

\address{%
  $^{1}$ Department of Computational Linguistics, University of Zurich, Switzerland. \\
  $^{2}$ Department of Informatics, University of Zurich, Switzerland. \\
  \{aref.farhadipour, srikanth.madikeriraghunathan\}@uzh.ch
}

\begin{document}

\maketitle
\begin{abstract}
    \sloppy
The CL-UZH team submitted one system each for the fixed and open conditions of the NIST SRE 2024 challenge. For the closed-set condition, results for the audio-only trials were achieved using the X-vector system developed with Kaldi. For the audio-visual results we used only models developed for the visual modality. Two sets of results were submitted for the open-set and closed-set conditions, one based on a pretrained model using the VoxBlink2 and VoxCeleb2 datasets. An Xvector-based model was trained from scratch using the CTS superset dataset for the closed set. In addition to the submission of the results of the SRE24 evaluation to the competition website, we talked about the performance of the proposed systems on the SRE24 evaluation in this report.
\end{abstract}

\section{Introduction}
Speaker recognition is an active research field that has witnessed significant progress in recent years, largely driven by advances in deep learning methods \cite{farhadipour2024analysis}. Despite these improvements, the task remains challenging in real-world conditions, where factors such as impaired or pathological speech \cite{farhadipour2024gammatonegram}, session variability, language mismatch \cite{zhang2025quantifying}, and short-duration utterances \cite{farhadipour2024analysis} can substantially degrade performance. To address these challenges, researchers are increasingly exploring the use of multimodal information, particularly the integration of audio and visual modalities \cite{farhadipour2024comparative}. Audio-visual speaker recognition leverages both speech signals and facial cues, offering a more robust representation of speaker identity. This multimodal approach has gained considerable attention in the community and is also a major focus in recent NIST Speaker Recognition Evaluations (SREs), highlighting its importance for the future of reliable and resilient speaker recognition systems.

We submitted the results for both closed-set and open-set conditions. The following subsections provide details of our submissions.

\section{Proposed System Architecture for SRE24 Challenge}

Our speaker verification system employs a multimodal approach, integrating both audio and visual cues to make robust decisions for each trial pair in the development and evaluation sets. The overall architecture, depicted in the figure above, comprises distinct processing streams for audio and visual modalities, followed by score fusion.

\begin{figure*}[t]
\centering
\includegraphics[width=0.9\linewidth]{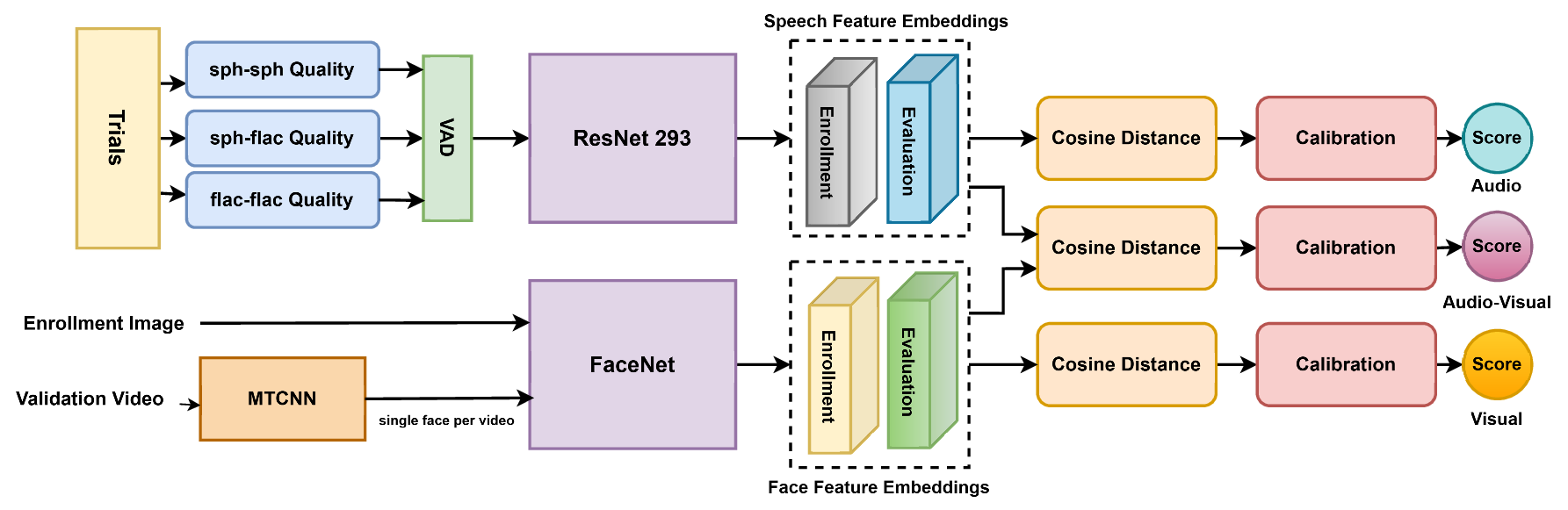}
\caption{Describe what the figure shows. For example: Architecture of the proposed audio-visual speaker verification system, illustrating the fusion of X-vector and FaceNet embeddings.}
\label{fig:diag3}
\end{figure*}
The audio processing pipeline begins with the Trials themselves, which are categorized based on the audio file types involved in each comparison: "sph-sph Quality," "sph-flac Quality," and "flac-flac Quality." This categorization allows for tailored processing and calibration based on the observed characteristics of these different trial types.

For enrollment and evaluation audio, a Voice Activity Detection (VAD) module is applied to filter out non-speech segments, ensuring that only relevant speech content is passed for further processing. The cleaned audio is then fed into a ResNet293 neural network, which extracts discriminative Speech Feature Embeddings. These embeddings are stored separately for "Enrollment" and "Evaluation" utterances.

For each trial, the cosine distance between the enrollment and evaluation speech embeddings is computed, yielding an initial similarity score. These scores then undergo a Calibration step, which is applied separately for each of the pre-defined trial quality categories (sph-sph, sph-flac, flac-flac). This modality-specific calibration further optimizes the reliability of the scores. The output of this stage is the final Audio Score.

The visual processing stream focuses on facial information. For enrollment, an Enrollment Image is directly input into a FaceNet model. For validation, a **Validation Video** is processed by an MTCNN (Multi-task Cascaded Convolutional Networks) algorithm, which detects faces and extracts a "single face per video" to ensure consistent input for subsequent steps.

The detected faces (from enrollment images and validation videos) are then processed by the FaceNet model, generating Face Feature Embeddings. Similar to the audio stream, these embeddings are maintained separately for "Enrollment" and "Evaluation" purposes.

The similarity between enrollment and evaluation face embeddings is quantified using Cosine Distance. These visual similarity scores are then passed through a dedicated Calibration module, resulting in the final Visual Score.

For trials involving both audio and visual data (Audio-Visual), the system combines information from both modalities. While not explicitly shown as a separate fusion block for a final AV score, the diagram implies that the individual Audio Scores and Visual Scores can be combined, potentially at the embedding level before cosine distance or at the score level, to derive a comprehensive decision. The arrow from "Speech Feature Embeddings" and "Face Feature Embeddings" to a unified "Audio-Visual" score indicates a potential concatenation of embeddings prior to cosine distance and calibration for the AV condition. This combined approach leverages the complementary strengths of both modalities, aiming to improve overall speaker verification accuracy.

\subsection{Fixed Condition}
For the fixed condition, we developed the X-vector~\cite{snyder2018x} based system with a Time-Delay Neural Network (TDNN)~\cite{peddinti2015time} backbone. Factorized TDNNs~\cite{povey2018semi} were used following ~\cite{madikeri2016idiap}.
The X-vector system was trained with the SRE21 CTS superset dataset.
Energy-based voice activity detection from Kaldi~\cite{kaldi_asru2011} was used in all data sets.
After VAD, approximately 6500 hours of speech data was retained with 6867 speakers.

PLDA classifier was trained on the SRE21 superset audio files~\cite{ioffe2006probabilistic}. Again, Kaldi was used
to train the PLDA classifier.
The scores were calibrated by training a linear transformation on the scores~\cite{alumae2019taltech}.
We used Kaldi~\cite{povey2011kaldi} toolkit for VAD, model training, embedding extraction, and scoring.

For the audio-visual condition, the audio was extracted from the video to obtain the embedding. The video segment was subsequently discarded.

The results on the SRE24 dev and eval sets are provided in Table~\ref{tab:closeset}. The NIST SRE 24 scoring scripts were used to generate the results. For speakers with multiple enrollments, the embeddings across audios are averaged.
In addition, we compared the performance of the X-vector system that takes AfV (Audio from Video) to considering only the video modality for recognition. For this experiment, we used the Facenet model (more details in Section \ref{tab:closeset}). Given that the performance of the X-vector system was significantly worse than that of the Facenet system, we submitted only the scores with the Facenet system.

\begin{table}[t]
\centering
\caption{Results on the SRE24 dev and eval sets for the fixed condition. The X-vector system was used for the audio condition. For the audio-visual condition, only the FaceNet model was used.}
\resizebox{0.45\textwidth}{!}{%  <-- Resizes table to text width; change \textwidth to e.g. 0.9\textwidth for 90%
\begin{tabular}{@{}lcccc|cccc@{}}
\toprule
\multirow{2}{*}{System} & \multirow{2}{*}{Condition} & \multicolumn{3}{c|}{Dev set} & \multicolumn{3}{c}{Eval set} \\
\cmidrule(lr){3-5} \cmidrule(lr){6-8}
& & EER(\%) & min\_C & act\_C & EER(\%) & min\_C & act\_C \\
\midrule
X-vector & Audio & 19.5 & 0.857 & 0.920 & 20.11 & 0.913 & 0.958 \\
FaceNet & AV & 13.5 & 0.440 & 0.606 & 17.64 & 0.491 & 0.699 \\
\midrule
FaceNet & Visual & 13.8 & 0.437 & 0.591 & -- & -- & -- \\
\bottomrule
\end{tabular}
}
\label{tab:closeset}

\end{table}

To compute the RTF of speaker embedding extraction, the embeddings were extracted on the SRE24 development set on a machine equipped with an Intel i9 CPU and 16 GB of RAM. The RTF achieved a value of 0.042. The RTF computation was performed in a single-threaded manner.

\subsection{Open Condition}

For the open condition, a system based on pre-trained neural networks was developed. Our system employs separate networks for audio and visual data, both trained on the Voxblink2 \cite{lin2024voxblink2} and VoxCeleb \cite{chung2018voxceleb2} datasets. Specifically, a ResNet293 network is used for audio processing and an ArcFace network \cite{deng2019arcface} is used for facial recognition.

For the audio component, given that the trial files involve four types of comparisons based on audio file type, individual speaker verification systems were designed for each case to optimize performance. The four types of systems arise from the four types of comparisons made based on the two audio file types: sph and flac. We observed that calibrating them based on the conditioned improved overall performance in terms of Equal Error Rate (EER).

For each speaker (model ID), only a single audio file is allocated for enrollment. WebRTC VAD~\footnote{\url{https://webrtc.org/}} is applied to sph files, while diarization data is utilized for FLAC files to remove extraneous segments. The processed audio files are then input to the ResNet293 network to extract embeddings, which are then compared with cosine distance scoring. The scores are subsequently calibrated, and a final score is derived. As mentioned earlier, the calibration is applied separately for each comparison type: sph vs sph, sph vs flac, flac vs sph and sph vs sph. In this part of the work, we used an NVIDIA 4090 GPU with 24 GB of VRAM, enabling processing at approximately 15 utterances per second. On the SRE 24 dev dataset, the RTF for embedding extraction was 0.00220.

For the visual module, a ResNet-based network utilizing ArcFace trained on Voxblink data, is employed. The strategy here is similar to the audio component: each speaker has a single enrollment sample, and only one frame is extracted for each image for testing. Frame extraction is conducted using the MTCNN algorithm \cite{7553523}. Embedding features from each image are then compared using cosine distance.

In the multimodal setup, the extracted vectors from each modality are concatenated, and cosine distance is applied to the combined vector to obtain the score for each trial.

The results for the open condition on both the SRE24 dev and eval sets are presented in Table~\ref{tab:openset}. The performance metrics, including EER, minDCF, and actDCF, are reported for each system and condition. The evaluation on the eval set provides a more robust assessment of the system's generalization capabilities beyond the development data.

\begin{table}[t]
    \centering
    \caption{Results on SRE24 dev set with the Resnet293 and Facenet models for open-set conditions conditions.}

    \resizebox{0.45\textwidth}{!}{%  <-- Resizes table to text width; change \textwidth to e.g. 0.9\textwidth for 90%
    \begin{tabular}{@{}lcccc|cccc@{}}
    \toprule
     \multirow{2}{*}{System} & \multirow{2}{*}{Condition} & \multicolumn{3}{c|}{Dev set} & \multicolumn{3}{c}{Eval set} \\ 
     \cmidrule(lr){3-5} \cmidrule(lr){6-8}
      & & EER(\%) & min\_C & act\_C & EER(\%) & min\_C & act\_C \\ 
     \midrule
     ResNet293 & Audio & 10.2 & 0.715 & 0.804 & 8.47 & 0.715 & 0.767 \\ 
     ResNet+FN & AV & 7.7 & 0.701 & 0.771 & 8.29 & 0.661 & 0.669 \\ 
     \midrule
     FaceNet & Visual & 13.82 & 0.437 & 0.591 & - & - & - \\ 
    \bottomrule
    \end{tabular}
    }
    \label{tab:openset}
    
\end{table}

\section{Conclusion}
This report detailed the CL-UZH team's submissions to the NIST SRE 2024 challenge for both fixed and open conditions. For the fixed condition, we presented an X-vector-based system for audio-only trials and a FaceNet-based approach for audio-visual and visual-only trials, demonstrating competitive performance on both dev and eval sets. For the open condition, we introduced a robust multimodal system integrating ResNet293 for audio and FaceNet (with ArcFace) for visual processing, trained on VoxBlink2 and VoxCeleb2. Our system effectively leveraged modality-specific calibration and achieved promising results, with the combined Audio-Visual system (ResNet+FN) consistently outperforming single-modality approaches, particularly evident in the lower EER values on both dev and eval sets. These results underscore the effectiveness of multimodal fusion and specialized network architectures in addressing the complexities of speaker recognition in challenging real-world scenarios.

\section{Acknowledgments}
Aref Farhadipour, Masoumeh Chapariniya, and Teodora Vukovic were funded by the CAPIRE project funded by DSI (Digital Society Initiative) at the University of Zurich.

\bibliographystyle{IEEEbib}
\bibliography{mystrings}

\begin{thebibliography}{10}

\bibitem{farhadipour2024analysis}
Aref Farhadipour and Hadi Veisi,
\newblock ``Analysis of deep generative model impact on feature extraction and dimension reduction for short utterance text-independent speaker verification,''
\newblock {\em Circuits, Systems, and Signal Processing}, vol. 43, no. 7, pp. 4547--4564, 2024.

\bibitem{farhadipour2024gammatonegram}
Aref Farhadipour and Hadi Veisi,
\newblock ``Gammatonegram representation for end-to-end dysarthric speech processing tasks: Speech recognition, speaker identification, and intelligibility assessment,''
\newblock {\em Iran Journal of Computer Science}, vol. 7, no. 2, pp. 311--324, 2024.

\bibitem{zhang2025quantifying}
Miao Zhang, Aref Farhadipour, Annie Baker, Jiachen Ma, Bogdan Pricop, and Eleanor Chodroff,
\newblock ``Quantifying and reducing speaker heterogeneity within the common voice corpus for phonetic analysis,''
\newblock {\em arXiv preprint arXiv:2506.00733}, 2025.

\bibitem{farhadipour2024comparative}
Aref Farhadipour, Masoumeh Chapariniya, Teodora Vukovic, and Volker Dellwo,
\newblock ``Comparative analysis of modality fusion approaches for audio-visual person identification and verification,''
\newblock {\em arXiv preprint arXiv:2409.00562}, 2024.

\bibitem{snyder2018x}
David Snyder, Daniel Garcia-Romero, Gregory Sell, Daniel Povey, and Sanjeev Khudanpur,
\newblock ``{X}-vectors: Robust {DNN} embeddings for speaker recognition,''
\newblock {\em Submitted to ICASSP}, 2018.

\bibitem{peddinti2015time}
Vijayaditya Peddinti, Daniel Povey, and Sanjeev Khudanpur,
\newblock ``A time delay neural network architecture for efficient modeling of long temporal contexts.,''
\newblock in {\em Interspeech}, 2015, pp. 3214--3218.

\bibitem{povey2018semi}
Daniel Povey, Gaofeng Cheng, Yiming Wang, Ke~Li, Hainan Xu, Mahsa Yarmohammadi, and Sanjeev Khudanpur,
\newblock ``Semi-orthogonal low-rank matrix factorization for deep neural networks.,''
\newblock in {\em Interspeech}, 2018, pp. 3743--3747.

\bibitem{madikeri2016idiap}
Srikanth Madikeri, Subhadeep Dey, Marc Ferras, Petr Motlicek, and Ivan Himawan,
\newblock ``Idiap submission to the nist sre 2016 speaker recognition evaluation,''
\newblock Tech. {R}ep., Idiap, 2016.

\bibitem{kaldi_asru2011}
{Daniel Povey et al.},
\newblock ``{The {K}aldi speech recognition toolkit},''
\newblock in {\em {Automatic {S}peech {R}ecognition and {U}nderstanding}}, 2011.

\bibitem{ioffe2006probabilistic}
Sergey Ioffe,
\newblock ``Probabilistic linear discriminant analysis,''
\newblock in {\em Computer Vision--ECCV 2006}, pp. 531--542. Springer, 2006.

\bibitem{alumae2019taltech}
Asadullah Tanel~Alum\"{a}e,
\newblock ``The {TalTech} systems for the {VOiCES from a Distance Challenge},''
\newblock in {\em Interspeech (submitted)}, 2019.

\bibitem{povey2011kaldi}
Daniel Povey, Arnab Ghoshal, Gilles Boulianne, Lukas Burget, Ondrej Glembek, Nagendra Goel, Mirko Hannemann, Petr Motlicek, Yanmin Qian, Petr Schwarz, et~al.,
\newblock ``The kaldi speech recognition toolkit,''
\newblock in {\em IEEE 2011 workshop on automatic speech recognition and understanding}. IEEE Signal Processing Society, 2011, number EPFL-CONF-192584.

\bibitem{lin2024voxblink2}
Yuke Lin, Ming Cheng, Fulin Zhang, Yingying Gao, Shilei Zhang, and Ming Li,
\newblock ``Voxblink2: A 100k+ speaker recognition corpus and the open-set speaker-identification benchmark,''
\newblock {\em arXiv preprint arXiv:2407.11510}, 2024.

\bibitem{chung2018voxceleb2}
Joon~Son Chung, Arsha Nagrani, and Andrew Zisserman,
\newblock ``Voxceleb2: Deep speaker recognition,''
\newblock {\em arXiv preprint arXiv:1806.05622}, 2018.

\bibitem{deng2019arcface}
Jiankang Deng, Jia Guo, Niannan Xue, and Stefanos Zafeiriou,
\newblock ``Arcface: Additive angular margin loss for deep face recognition,''
\newblock in {\em Proceedings of the IEEE/CVF conference on computer vision and pattern recognition}, 2019, pp. 4690--4699.

\bibitem{7553523}
K.~Zhang, Z.~Zhang, Z.~Li, and Y.~Qiao,
\newblock ``Joint face detection and alignment using multitask cascaded convolutional networks,''
\newblock {\em IEEE Signal Processing Letters}, vol. 23, no. 10, pp. 1499--1503, Oct 2016.

\end{thebibliography}

\end{document}